\documentclass[preprint,12pt]{aastex}

\def\msun{{\rm\,M_\odot}}

\newcommand{\etal}{et al.\ }

\def\h2{${\rm\,H_2}$}

\slugcomment{Submitted to ApJ}

\begin{document}

\title{Implications of WMAP Observations On the Population III Star Formation Processes}

\author{Renyue Cen\altaffilmark{1}}

\altaffiltext{1} {Princeton University Observatory, 
Princeton University, Princeton, NJ 08544; cen@astro.princeton.edu}

\accepted{ }

\begin{abstract}

In an earlier paper (Cen 2003) we pointed out the strong likelihood
for the universal reionization to occur twice, giving rise to a 
much larger Thomson optical depth due to 
the intergalactic medium than that in the case
of a single rapid reionization at $z\sim 6$.
The latest Wilkinson Microwave Anisotropy Probe (WMAP) observations
(Kogut \etal 2003)
indicate that the universe indeed appears to have entered
a significantly ionized state at a very high redshift.
In light of this new development, 
we perform a more focused analysis of the 
Thomson optical depth in the context of the spatially flat,
cosmological constant-dominated cold dark matter
model constrained by WMAP observations.

While the current uncertainties on the observed 
Thomson optical depth are still relatively large, with 
$\tau_e=0.17\pm 0.04$ ($68\%$) (Kogut \etal 2003),
important implications on Pop III star formation processes
at high redshift can already be inferred.
We are able to draw four conclusions:
(1) in the absence of a top-heavy initial stellar mass function (IMF)
for Pop III metal-free stars and without a dramatic
upturn in the star formation efficiency and ionizing photon escape
fraction at high redshift ($z>6$), we find 
$\tau_e \le 0.09$;
(2) with a top-heavy IMF for the Pop III metal-free stars
and plausible star formation efficiency and ionizing photon escape fraction,
it is expected that $\tau_e \le 0.12$;
(3) it is possible to reach $\tau_e = 0.15$,
if the metal enrichment efficiency of the intergalactic medium 
by Pop III stars is very low thus Pop III era is prolonged;
(4) to reach $\tau_e \ge 0.17$ requires
either of the following two conditions:
the cosmological model power index $n$ is positively tilted to $n\ge 1.03$,
Pop III star formation in minihalos with molecular hydrogen cooling
has an efficiency $c_*(H_2,III)>0.01$ (with ionizing photon
escape fraction greater than 30\%).
Thus, if the current observed value of 
Thomson optical depth withstands future data,
we will have strong observational evidence
that Pop III stars are massive and their formation efficiency 
may be much higher than current theoretical works suggest.
Alternatively, there may be 
unknown, non-stellar ionizing sources at very high redshift.

\end{abstract}


\section{Introduction}

The cosmological reionization process is likely to be quite complex 
(e.g., Barkana \& Loeb 2001; Madau 2002;
Wyithe \& Loeb 2003a;
Mackey, Bromm, \& Hernquist 2002;
Cen 2003;
Venkatesan, Tumlinson, \& Shull 2003).
It seems rather remarkable that we can now start observationally probing 
this process.
A major milestone was laid down by 
the recent observations of high redshift 
quasars from the Sloan Digital Sky Survey (SDSS),
indicating that the final reionization episode 
came to completion at $z\sim 6$
(e.g., Fan \etal 2001; Becker \etal 2001; Barkana 2002; 
Cen \& McDonald 2002; Litz \etal 2002).
Given the final reionization epoch
and the fact that the density fluctuations at that epoch
is sufficiently well constrained in the context of the standard
cold dark matter model (Bahcall \etal 1999),
it becomes possible to show (Cen 2003), under reasonable assumptions,
that the universe has been reionized twice,
first reionized at $z\ge 15$ by
metal-free, massive Population III (Pop III) stars 
and second reionized at $z=6$,
{\it provided that Pop III IMF is top-heavy}.
In this picture a larger Thomson optical depth,
$\tau_e \ge 0.07$, was expected.

Another major milestone was set very recently by the
WMAP polarization observations,
which probes the ionization state of the gas in the universe
at high redshift during the cosmological reionization period.
The WMAP observations measured 
$\tau_e=0.17\pm 0.04$ ($68\%$) (Kogut \etal 2003).
One can immediately draw the conclusion 
that the universe did not just experience 
an impulsive reionization event at $z\sim 6$,
which would have yielded $\tau_e\sim 0.03$,
in the standard cosmological constant dominated
cold dark matter model.
Evidently, the universe became (and possibly had to stay)
significantly ionized
from a much higher redshift, probably from $z\ge 15$.
In this {\it Letter} we compute the detailed reionization
process and compare to WMAP observations,
investigating what can be learned about
Pop III star formation processes at high redshift.

\section{Thomson Optical Depth $\tau_e$}

A spatially flat cold dark matter cosmological model 
with $\Omega_M=0.27$, $\Omega_b=0.047$, $\Lambda=0.73$,
$H_0=72$km/s/Mpc, $n_s=0.99$ and $\sigma_8=0.90$,
symbolized as LCDMAP model, is used.
This model is the best fit model constrained by WMAP observations
with a fixed power-law index (Spergel \etal 2003).
To set a plausible upper bound we also
compute a model with $n_s=1.03$ and $\sigma_8=1.0$,
denoted as LCDMAP+ model,
which is consistent with current WMAP data
(Spergel \etal 2002) and maximizes the small-scale
structures responsible for the reionization process.
We employ an efficient method 
to compute the coupled evolution of gas and star formation
in the high redshift universe, described in \S 4 of Cen (2003),
which simultaneously computes the dynamic evolution 
of all gaseous phases
including HII regions, HI regions and partially ionized regions.
Such a careful treatment is necessary given the various short
time scales involved at high redshift for important
processes, including recombination and cooling.
Throughout, Pop III stars are defined to be metal-free
stars with a top-heavy initial mass function (IMF)
and Pop III galaxies are defined to be galaxies 
where Pop III stars form.
We define ``minihalos" as those whose
virial temperature is less than $\sim 1\times 10^4$K where
only \h2 cooling is possible in the absence of metals,
and large halos as those with 
virial temperature above $\sim 1\times 10^4$K
capable of cooling via atomic lines.

The most important factors determining
the ionization process are
$\Psi\equiv c_*\times f_{\rm esc}$ and $C$,
where $c_*$ is the star formation efficiency,
defined as the fraction of gas in halos formed into stars;
$f_{\rm esc}$ is the fraction of ionizing photons
that escapes from galaxy halos and gets into the IGM;
$C$ is the clumping factor of the intergalactic medium (IGM).
Basically, $\Psi$ is the source of ionizing
photons from stars in galaxies and 
$C$ determines the sink of ionizing photons due to the IGM.
The competition between the two primarily dictates
the ionization state of the IGM.
The initial phase of ionization front propagating
through neutral medium
does not play a significant role near the 
end of the reionization process,
and cosmological effects 
are unimportant throughout the reionization period.
But all these processes are included in the calculation.
Theoretically, $C$ is reasonably well determined,
if we adopt the standard cold dark matter cosmological model.
In addition, $C$ may be constrained directly by observations
of quasar absorption spectra at redshifts
close to $z=6$ (e.g., Cen \& McDonald 2002; Djorgovski \etal 2001;
Cen \& Haiman 2000).
We set the IGM clumping factor at $z=6$ to be
$50.0$, according to Gnedin \& Ostriker (1997),
which requires $C_{\rm halo}=702$ with the adopted cosmological model
(see Cen 2003 for discussion).
Results do not sensitively depend on $C$ at $z=6$;
for example, a change in $C$ at $z=6$ by $25\%$ only causes 
$\sim 5\%$ change in $\tau_e$.
Both theoretically and observationally, 
our direct knowledge of $\Psi$ for high redshift
galaxies is very little.
However, as pointed out in Cen (2003),
the fact that the universal reionization 
ends just at $z\sim 6$, as observations seem to indicate
(e.g., Fan \etal 2001; Becker \etal 2001; Barkana 2002; 
Cen \& McDonald 2002; Litz \etal 2002),
tightly constrains $\Psi$, knowing $C$.
Such a normalization point is proven to be very powerful
and unambiguously determines $\Psi$
at $z\sim 6$ (see Figure 7 of Cen 2003) 
for Pop II stars,
which are likely responsible
for completing the long reionization process at $z\sim 6$.
For all the models computed below, the normalization at $z=6$
requires $\Psi(II)\equiv c_*(II)\times f_{esc}(II)=0.015-0.017$
for Pop II galaxies at 
$z\sim 6$.
For models computed we assume $\Psi(II)$ to be constant with redshift,
except indicated otherwise.
On the other hand, $c_*(III)$ and $f_{esc}(III)$ are less constrained.

\begin{deluxetable}{cccccccccc} 
\tablewidth{0pt}
\tablenum{1}
\tablecolumns{6}
\tablecaption{Summary of Star Formation Models} 
\tablehead{
\colhead{\#} &
\colhead{Model} &
\colhead{$c_*$({\rm H$_2$, III})} &
\colhead{$c_*$({\rm HI, III})} &
\colhead{$f_{crit}$} &
\colhead{$f_{\rm esc}(III)$} &
\colhead{$\tau_e$} }

\startdata
{\rm 1} & LCDMAP & 0 & 0 & $0$ & - & 0.07 \nl 
{\rm 2} & LCDMAP+ & 0 & 0 & $0$ & - & 0.08 \nl 
{\rm 3} & LCDMAP & 0.002 & 0.1 & $1\times 10^{-4}$ & (0.05,0.1,0.2,0.3) & (0.09,0.10,0.12,0.13) \nl 
{\rm 4} & LCDMAP+ & 0.002 & 0.1 & $1\times 10^{-4}$ & (0.05,0.1,0.2,0.3) & (0.10,0.12,0.14,0.16) \nl 
{\rm 5} & LCDMAP & 0.01 & 0.1 & $1\times 10^{-4}$ & (0.05,0.1,0.2,0.3) & (0.09,0.11,0.14,0.15) \nl 
{\rm 6} & LCDMAP+ & 0.01 & 0.1 & $1\times 10^{-4}$ & (0.05,0.1,0.2,0.3) & (0.10,0.13,0.16,0.19) \nl 
{\rm 7} & LCDMAP & 0.002 & 0.1 & $5\times 10^{-4}$ & (0.05,0.1,0.2,0.3) & (0.11,0.12,0.14,0.15) \nl 
\enddata
\end{deluxetable}

We compute seven models with
varying parameters for Pop III star formation processes
and cosmological parameters, listed in Table 1,
to show the possible range 
for the Thomson optical depth due to the IGM at reionization.
In Table 1, 
$c_*(H_2,III)$ indicates the star formation
efficiency in minihalos with virial temperature
less than $8,000$K, where \h2 cooling dominates for metal-free gas;
$c_*(HI,III)$ indicates the star formation
efficiency in large halos with more efficient atomic cooling;
$f_{crit}$ indicates the fractional amount of gas formed
into Pop III stars at the transitional epoch from Pop III to Pop II
(which is assumed to determine the metallicity of the 
IGM; see Cen 2003);
$f_{esc}(III)$ is the mean fraction of ionizing photons
produced by galaxies that escapes from the halos
of Pop III galaxies and into the IGM;
the last column $\tau_e$ is the integrated (from $z=0$ to $z=1000$)
Thomson optical depth due to the IGM,
where multiple entries correspond to 
respective multiple entries for $f_{esc}(III)$.

Models \#1,2 represents the case where
the IMF at high redshift is assumed to remain as the Salpeter (1955)
function with the same low-mass cut-off ($0.1\msun$), as at $z=6$,
which is the likely minimum model in terms of $\tau_e$,
for LCDMAP and LCDMAP+ model, respectively.
Models \#3,4 have a Pop III star formation efficiency
in minihalos $c_*(H_2,III)=0.002$,
suggested by numerical simulations of Abel, Bryan \& Norman (2002)
with a reasonable star formation efficiency for Pop III
galaxies with large halos of $c_*(HI,III)=0.10$,
for LCDMAP and LCDMAP+ model, respectively.
Models \#5,6 assume
a higher Pop III star formation efficiency
in minihalos $c_*(H_2,III)=0.01$,
for LCDMAP and LCDMAP+ model, respectively;
in this case, 
ionizing photons from Pop III minihalos 
dominate over those from Pop III large halos
for the first reionization process,
regardless of the value of $c_*(HI,III)$.
Model \#7 is similar to Models \#1,
except that in the former we arbitrarily set
the transition epoch from Pop III to Pop II
at the time when a fraction $f_{\rm crit}=5\times 10^{-4}$ of total
gas has formed Pop III stars instead of 
$f_{\rm crit}=1\times 10^{-4}$ in the latter set.
We note that $f_{\rm crit}=1\times 10^{-4}$ 
is derived based on the assumption that most of the 
Pop III metal-free massive stars experience supernovae and 
most of the ejected metals get out of galaxies
to enrich the IGM (Oh \etal 2001; Cen 2003).
If, for some reason,
the fractional amount of metals enriching the IGM
is significantly smaller than the amount of 
gas formed into massive stars,
the efficiency of metal enrichment by Pop III stars
would be lower; Model \#7 serves to illustrate this possibility.
Possible physical processes that may reduce the metal enrichment
efficiency include 
a large fraction of Pop III stars being very massive stars 
($M\ge 300\msun$) which may collapse to black holes
instead of exploding as supernovae 
(Rakavy, Shaviv, \& Zinamon 1967;
Bond, Arnett, \& Carr 1984; 
Glatzel, Fricke, \& El Eid 1985; Woosley 1986)
or not all ejected metals  are able to be transported to IGM.
We note that Model \#1 is close
to the model used in Cen (2003),
having perhaps more conventional (thus thought to be more reasonable)
values for the relevant parameters.
We also test a variant of Model \#1 by using
$\Psi(II)\propto (1+z)^{1/2}$ instead of being constant with redshift
and find $\tau_e=0.08$ for that case versus $0.07$ for
a constant $\Psi(II)$ (Model \#1).

From Table 1 we can draw several conclusions.
First, without Pop III stars (Models \#1,2),
it is likely that $\tau_e < 0.09$ for both LCDMAP and LCDMAP+ models,
inconsistent with WMAP results at $\ge 2\sigma$ level (Kogut \etal 2003).
The conclusions reached here
with regard to the requirement of a top-heavy IMF 
for Pop III stars in order to explain
the observed high $\tau_e$ value
are consistent with those by Wyithe \& Loeb (2003b),
Haiman \& Holder (2003) and Sokasian \etal (2003).
Second, for reasonable ranges for the star formation efficiency
[$c_*(HI,III)\le 0.1$, $c_*(H_2,III)\le 0.002$]
and ionizing photon escape fraction ($f_{esc}(III)\le 0.20$)
from Pop III galaxies,
we expect that $\tau_e \le 0.12$ for LCDMAP model (Model \#3)
and $\tau_e \le 0.15$ for LCDMAP+ model (Model \#4).
Third, prolonging the Pop III era by making
the metal enrichment of the IGM from Pop III galaxies
less efficient increases the Thomson optical depth incrementally.
Finally, 
while $\tau_e=0.17$ is the mean value determined by WMAP observations,
it seems hard to reach 
in the LCDMAP model.
There are two possible ways to achieve a Thomson optical
depth as high as $\tau_e\ge 0.17$:
either the ionizing photon escape fraction
is large ($f_{esc}(III)\ge 0.3$) and the underlying cosmological model
has a blue power spectrum with a positive tilt
to $n\ge 1.03$ (Model \#5) 
or Pop III star formation efficiency in minihalos
is substantially larger than current simulations seem to indicate,
requiring $c_*(H_2,III)>0.01$.

Figures 1,2 show the reionization history and
cumulative Thomson optical depth, respectively,
for four representative models listed in Table 1.
We see that without Pop III stars (Model \#1)
reionization event only occurs once at $z\sim 6$
and the reionization history is monotonic.
With Pop III stars 
the first reionization occurs at $z=15-20$ (Models \#3,4).
In the last case (Model \#5) where 
Pop III star formation efficiency 
in minihalos is substantially larger than 
suggested by current simulations of Abel, Bryan \& Norman (2002),
$0.01$ versus $\sim 0.002$,
the first reionization occurs at $z\sim 20-25$.
While both Model \#4 and Model \#5 have comparable 
total Thomson optical depth, $\tau_e=0.15-0.16$,
their respective reionization histories at early
times ($z\ge 15$) are somewhat different.
We find that for Models \#4,
minihalos and large halos contribute
to 62\% and 38\% of all ionizing photons to 
reionize the universe at the first time,
by which 9 ionizing photons per baryon have been produced;
the corresponding numbers for Model \#5
are 61\%, 39\% and 11 photons per baryon.
In contrast, for Model \#3, 
minihalos and large halos contribute
to 20\% and 80\% of all ionizing photons to 
the first reionization and 
6 ionizing photons per baryon have been produced by then.
We see, as expected,
that an earlier first reionization requires a larger
number of ionizing photons per baryons
and a larger contribution from minihalos appears 
likely if the first reionization occurs at $z\ge 20$
in the context of the currently favored $\Lambda$ cosmology.
It is possible that
future CMB experiments such as Planck survey
might be able to distinguish between these two scenarios
(e.g., Holder \etal 2003).
We will defer a more detailed analysis of this aspect
to a separate paper.

\begin{figure}
\plotone{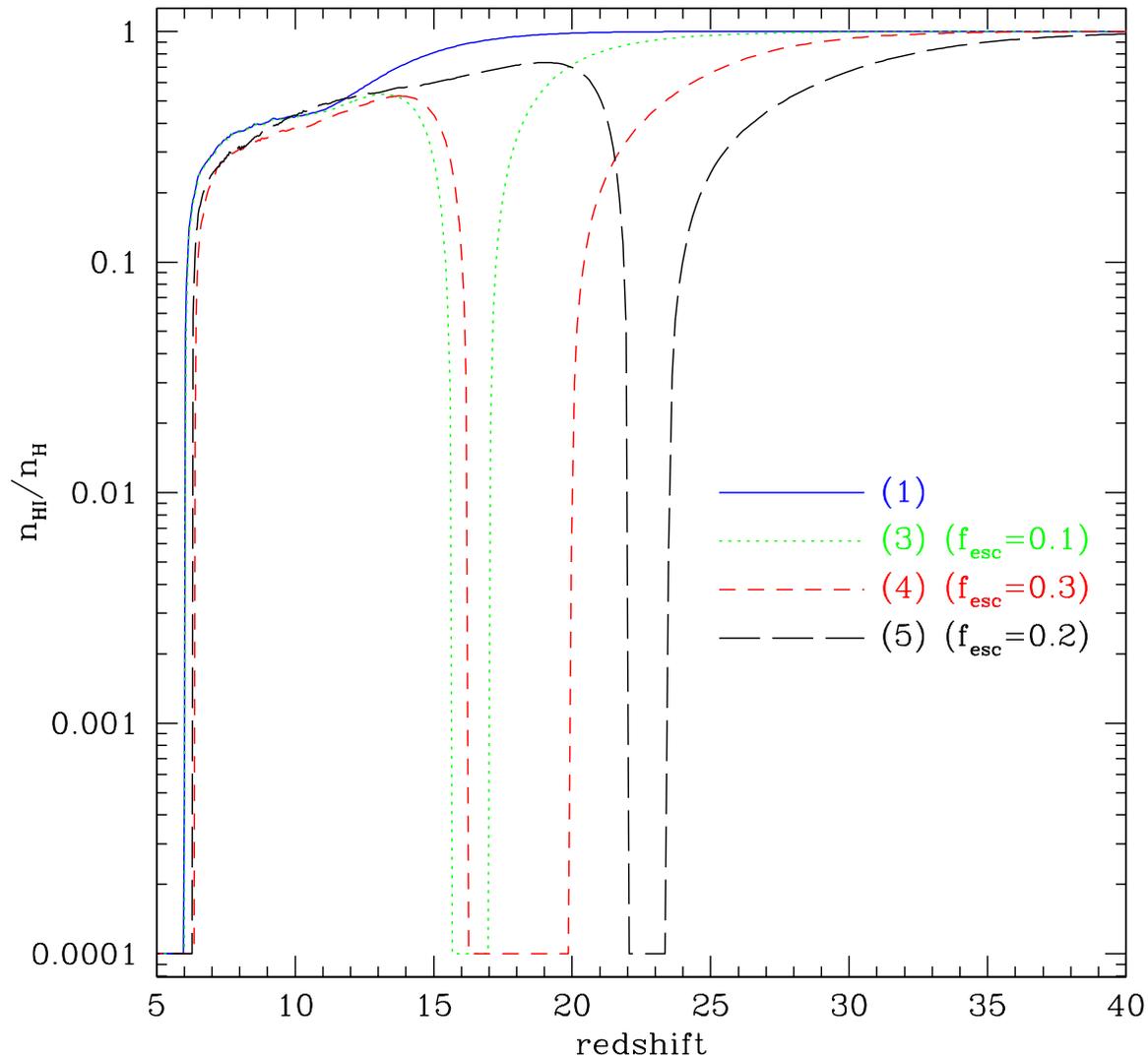}
\caption{
shows the global mean of 
the hydrogen neutral fraction as a function of redshift
for four models 
(Model \#1 : solid, blue;
Model \#3 with $f_{esc}(III)=0.1$: dotted, green;
Model \#4 with $f_{esc}(III)=0.3$: short dashed, red;
Model \#5 with $f_{esc}(III)=0.2$: long dashed, black)
in Table 1.
}
\label{y}
\end{figure}

\begin{figure}
\plotone{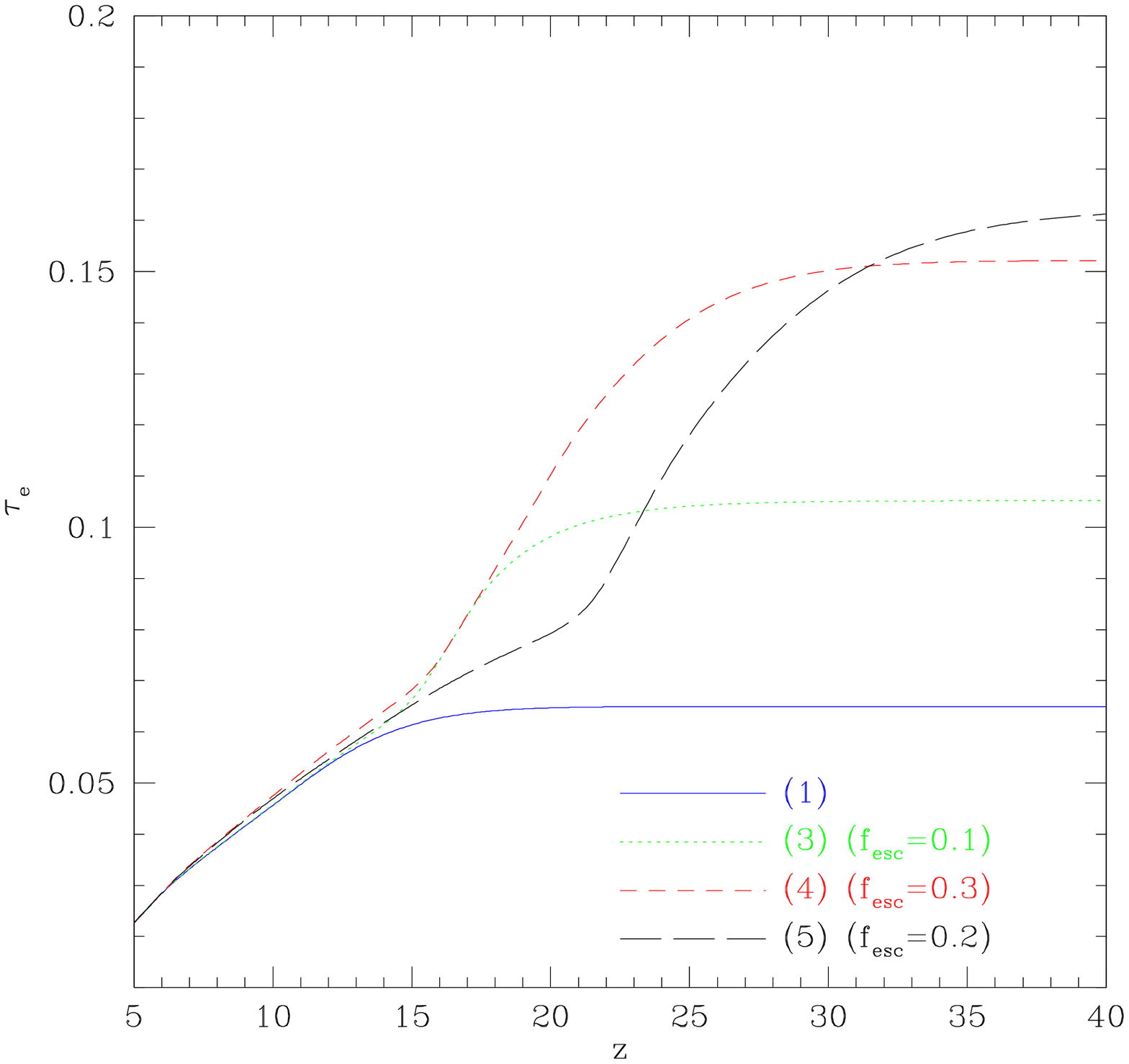}
\caption{
shows the cumulative Thomson scattering 
optical depth as a function of redshift
for four models 
(Model \#1 : solid, blue;
Model \#3 with $f_{esc}(III)=0.1$: dotted, green;
Model \#4 with $f_{esc}(III)=0.3$: short dashed, red;
Model \#5 with $f_{esc}(III)=0.2$: long dashed, black)
in Table 1.
}
\label{tau}
\end{figure}

\section{Discussion}

Since it appears that a significant fraction 
($f_{esc}(III)\ge 0.1$) of ionizing
photons produced by Pop III galaxies is required to escape
from galaxies in order to sufficiently ionize
the universe to be consistent with WMAP observations,
it is worthwhile to check what kind of escape fraction is possible.
Here we give an order of magnitude assessment of the situation.
If a fraction,
$c_*({\rm III})$, of gas in a high redshift
Pop III galaxy formed into very massive stars
emitting at $1.6\times 10^{48}$ hydrogen
ionizing photons per solar mass of stars per second (e.g.,
Bromm, Kudritzki, \& Loeb 2001),
then we can obtain the ratio of the total number of hydrogen ionizing 
photons emitted during the $3\times 10^6$ yrs
of lifetime of the Pop III galaxy (assumed to be equal
to the lifetime of the Pop III massive stars) 
to the number of baryons in the galaxy,
\begin{equation}
{N_{\rm ph}\over N_{\rm H}} = 1.3\times 10^5 c_*({\rm III}).
\end{equation}
\noindent
It can also be shown that the 
ratio of the number of recombinations of hydrogen atoms 
(assuming a temperature of $1.5\times 10^4~$K at $z=20$)
to the number of baryons in the galaxy is 
\begin{equation}
{N_{\rm rec}\over N_{\rm H}} = 0.061 C_{\rm g},
\end{equation}
\noindent
where $C_{\rm g}$ is the effective clumping factor
of the gas in the galaxy, taking into account all possible effects
including shielding and clumpiness.
Thus, the ratio of 
the total number of ionizing photons
to the total number of recombinations of hydrogen atoms 
during the lifetime of the galaxy at redshift $z=20$ is 
\begin{equation}
{N_{\rm ph}\over N_{\rm rec}} = 2.1\times 10^6 C_{\rm g}^{-1} c_*({\rm III}).
\end{equation}
\noindent
The time it takes for the ionization front from very massive Pop III stars 
to break out is roughly
\begin{equation}
t_{\rm IF} = {4\pi r^3 n\over L_{\rm ph}} = 6.8 ({M_*\over 100\msun})^{-1} ({r\over 1{\rm pc}})^3 ({n\over 10^4 {\rm cm}^{-3}})~{\rm yrs},
\end{equation}
\noindent
which is clearly much shorter than the lifetime
of the galaxy. 
In Equation (4) $M_*$ is the amount of Pop III stars
formed, $r$ the size of the galaxy disk and $n$ the interstellar
medium density.
Thus, ionization of the gas
inside the galaxy may be treated as instantaneous
and the ratio in Equation (3) is an instructive gauge.

If $c_*(HI, {\rm III})\ge 0.1$, as appears to be required
(as discussed in \S 2)
in order to produce a large enough
$\tau_e$ ($\ge 0.1$)
to be consistent with WMAP observations,
then Equation (3) implies that $C_g$ has to be 
less than $2.1\times 10^5$ in order for
a significant fraction of ionizing photons to escape
from galaxy halos.
On one hand, if the clumping factor 
is dominated by gas in the halo with 
overdensity of order $100$, $C_g$ could
be less than $2.1\times 10^5$.
On the other hand, the gas on the disk or at radii
much smaller than the virial radius 
would be much denser and their contribution
to the clumping factor would critically
depend on the density, geometry and self-shielding effects.
A reliable estimate of the escape fraction for ionizing
photons is unattainable without detailed modeling.
Nevertheless, it appears that
it is not entirely impossible to have
a significant fraction of ionizing photons escape 
under special conditions.
Possible favorable scenarios include a large fraction 
of star formation activities occur 
at a substantial height above the galactic disk and/or off-center
within star formation sites embedded in very clumpy interstellar medium.

It is clearly seen from the above analysis
that the smaller the value of $c_*({\rm III})$,
the less likely that a significant fraction of ionizing 
photon can escape.
Consequently,
a high Pop III star formation efficiency 
may be the key to a high overall ionizing photon production rate
in Pop III galaxies with respect to the receiving IGM,
with the latter depending on the former in a probably nonlinear way.
The question is then:
how high can $c_*({\rm III})$ be?

The simulation by Abel, Bryan \& Norman (2002; see 
also Bromm, Coppi, \& Larson 2002)
for a halo of mass $10^6\msun$ with $6\times 10^4\msun$ baryons
forms a massive star of mass $100\msun$ at the center.
This would give a star formation efficiency of
$0.0017$, taken at face value, 
which is much smaller
than required value of order $0.01$.
For a discussion why Pop III star formation in minihalos 
may continue unimpeded throughout the Pop III era, see Cen (2003).
The value of $0.1$ required for Pop III stars
in large halos to ionize the universe
at $z=15-20$ is still higher by a factor of ten.
It should be pointed out that
that the central protostar in the Abel \etal (2002) simulation
appears to be still accreting matter rapidly at
the end of their simulation,
so it is possible that the stellar mass may grow substantially.
We also note, however, that their simulated galaxy
has a virial temperature below $1\times 10^4~$K hence
does not possess the efficient atomic cooling.
It is possible that, for galaxies with efficient
atomic cooling, 
star formation efficiency may be much higher than 
what Abel \etal (2002) simulation indicates.
Detailed simulations will be invaluable.

Finally, we note that,
if star formation in cooling shells produced by exploding
massive Pop III stars,
the ``Pop II.5" stars proposed by Mackey \etal (2003),
is efficient,
they may make a non-negligible contribution
to the pool of ionizing photons.
These stars might form at off-center locations
and thus might possess a relatively larger ionizing photon
escape fraction.
The counteracting factor is that 
these stars might be substantially less massive 
than the first generation metal-free stars,
since the metallicity of the gas in the cooling shells
may be quite high,
hence are less efficient ionizing photon emitters.

\section{Conclusions}

Adopting the best fit standard cold dark matter model
with a fixed power-law index by WMAP observations 
with $\Omega_M=0.27$, $\Omega_b=0.047$, $\Lambda=0.73$,
$H_0=72$km/s/Mpc, $n_s=0.99$ and $\sigma_8=0.90$ 
(Spergel \etal 2003)
and based on the observed Thomson optical depth due to 
intergalactic medium by WMAP (Kogut \etal 2003),
we are able to draw several relatively secure conclusions
with regard to Pop III star formation
processes at very high redshift.

(1) The combination 
of the normal Salpeter IMF for Pop III metal-free stars and 
the absence of a dramatic
upturn in the star formation efficiency and/or ionizing photon escape
fraction at high redshift ($z>6$)
would produce a Thomson optical
depth due to IGM at reionization of $\tau_e \le 0.09$,
inconsistent with 
the observed $\tau_e=0.17\pm 0.04$ ($68\%$) (Kogut \etal 2003)
at $\ge 2\sigma$ level.

(2) A top-heavy IMF for the Pop III metal-free stars
and plausible star formation efficiency and ionizing photon escape,
as gauged by the corresponding values for Pop II galaxies
required in order to achieve the second
reionization finale at $z\sim 6$ yield $\tau_e \le 0.12$;

(3) In the event that
the metal enrichment efficiency of the intergalactic medium 
by Pop III stars is very low thus Pop III era is prolonged,
one may be able to obtain $\tau_e = 0.15$.

(4) It seems quite improbable to reach $\tau_e \ge 0.17$
even with very massive Pop III metal-free stars,
unless (i)
the cosmological model power index $n$ is positively tilted to $n\ge 1.03$
and/or (ii) Pop III star formation in minihalos with molecular hydrogen cooling
has an efficiency $c_*(H_2,III)>0.01$ (still requiring ionizing photon
escape fraction greater than $0.3$)
or (iii) alternatively, there may be unknown, non-stellar 
ionizing sources at very high redshift
in the context of the $\Lambda$CDM cosmology.

It is expected that the errorbars 
on the detected Thomson optical depth
should shrink significantly with time using more data.
Thus, if the current observed value of 
Thomson optical depth ($\tau_e\ge 0.09$ - currently $2\sigma$ lower
bound) withstands future data,
we will have strong observational evidence
for the existence of massive metal-free Pop III stars.
One possible outcome will be that
$0.09\le\tau_e\le 0.12$ ($1.25-2.0\sigma$ away from the current observed
central value), in which case no drastic increase in the star formation
efficiency in Pop III galaxies would be needed.
On the other hand, if $\tau_e\ge 0.17$ is firmed up in the future,
there will be a degeneracy between possible competing scenarios.
Future CMB experiments such as the Planck surveyor
may be able to distinguish between them.
On the theoretical side, more accurate
calculations of Pop III metal-free star formation
in both minihalos and large halos will be extremely
useful to shed light on these vital parameters.

\acknowledgments
I thank Zoltan Haiman, Lars Hernquist,
Eve Ostriker, David Spergel and Jonathan Tan 
for useful discussion.
This research is supported in part by NSF grant AST-0206299.

\end{document}